\documentclass{aa}
\usepackage{graphicx,natbib}
\usepackage[latin1]{inputenc}
\usepackage[T1]{fontenc}
\bibpunct{(}{)}{;}{a}{}{,}
\usepackage{txfonts}

\def\lj{L_{\rm J}}
\def\fgas{f_{\rm gas}}
\def\fvol{f_{\rm vol}}

\def\mmol{M_{\rm Mol}}
\def\tetad{\theta_{\rm D}}

\def\mearth{M_{\oplus}}

\def\rearth{R_{\oplus}}

\def\f1{f_{\rm I}}

\def\beq{\begin{equation}}
\def\eeq{\end{equation}}

\def\mearth{M_\oplus}

\def\mearth{M_\oplus}

\def\simgr{\,\hbox{\hbox{$ > $}\kern -0.8em \lower 1.0ex\hbox{$\sim$}}\,}
\def\simle{\,\hbox{\hbox{$ < $}\kern -0.8em \lower 1.0ex\hbox{$\sim$}}\,}
\def\beq{\begin{equation}}
\def\eeq{\end{equation}}

\def\simgr{\,\hbox{\hbox{$ > $}\kern -0.8em \lower 1.0ex\hbox{$\sim$}}\,}
\def\simle{\,\hbox{\hbox{$ < $}\kern -0.8em \lower 1.0ex\hbox{$\sim$}}\,}
\def\beq{\begin{equation}}
\def\eeq{\end{equation}}

\def\apj{ApJ}                 
\def\apjl{ApJ}                
\def\apjs{ApJS}               
\def\ao{Appl.Optics}          
\def\aap{A\&A}                

\def\({\left(}
\def\){\right)}
\def\<{\left<}
\def\>{\right>}

\def\({\left(} 
\def\){\right)} 
\def\<{\left<} 
\def\>{\right>} 

\def\bc{\begin{changebar}}
\def\bce{\begin{center}}
\def\beq{\begin{equation}} 

\def\bi{\begin{itemize}}

\def\btab{\begin{tabular}{p{1.7cm}p{12cm}p{1.5cm}}}
\def\bt2{\begin{tabular}{p{1 cm}p{4.5cm}p{10cm}}}

\def\ec{\end{changebar}}
\def\ece{\end{center}}
\def\eeq{\end{equation}} 
\def\ei{\end{itemize}}

\def\etab{\end{tabular}\\}

\def\mH2{m_\mathrm{H_2}}

\def\dmax0{\rho_\mathrm{max}}
\def\dmaxS0{\Sigma_\mathrm{max}}

\def\rH2{r_\mathrm{H_2}}

\def\r0max{r_\mathrm{0max}}

\def\s0{\sigma_\mathrm{0}}

\def\xp0{x_{\rm{M0}}}

\def\z0max{z_\mathrm{0max}}

\def\apj{{\it ApJ}}                 
\def\apjl{ApJ}                
\def\apjs{ApJS}               
\def\ao{Appl.Optics}          
\def\aap{{\it A\&A}}                

\usepackage{changebar}
\usepackage{bm}
\usepackage{units}
\usepackage{color}

\begin{document}

\title{Constraining the volatile fraction of planets from transit observations}

\author{Y. Alibert \inst{1}}
\offprints{Y. Alibert}
\institute{Physikalisches Institut  \& Center for Space and Habitability, Universitaet Bern, CH-3012 Bern, Switzerland, 
        \email{yann.alibert@space.unibe.ch}
}

\abstract
{The determination of the abundance of volatiles in extrasolar planets is very important as it can provide constraints on transport in protoplanetary disks
and on the formation location of planets. However, constraining the internal structure of low-mass planets from transit measurements is known to be a  degenerate problem. }
{Using planetary structure and evolution models, we show how observations of transiting planets can be used to constrain their internal composition,
in particular the amount of volatiles in the planetary interior, and consequently the amount of gas (defined in this paper to be only H and He) that the planet harbors.
 {We first explore planets that are located close enough to their star to have lost their gas envelope. We then
concentrate on planets at larger distances and} show that the observation of transiting planets at different evolutionary ages can provide statistical information
on their internal composition, in particular on their volatile fraction.}
{We computed the evolution of low-mass planets (super-Earths to Neptune-like) for different fractions of volatiles and gas. We used
a four-layer model (core, silicate mantle, icy mantle, and gas envelope) and computed the internal structure of planets for different luminosities.
With this internal structure model, we computed the internal and gravitational energy of planets, which was then used to derive the time evolution of the planet.
Since the total energy of a planet depends on its heat capacity and density distribution and therefore on its composition, planets with different ice fractions have different
evolution tracks.}
{ {We show for low-mass gas-poor planets that are located close to their central star that \textit{\textup{as}\textup{suming evaporation has efficiently removed the entire gas envelope}}, it is possible to constrain the volatile fraction
of close-in transiting planets. We illustrate this method on
the example of 55 Cnc e and show that under the assumption of the absence of gas, the measured mass and radius imply 
 at least 20 \% of volatiles in the interior. For planets at larger distances,} we show that the observation of transiting planets at different evolutionary ages can be used to set statistical constraints on the volatile content of planets.}
{These results can be used in the context of future missions like PLATO to better understand the internal composition of planets, and based on this,
their formation process and potential habitability.} 

\keywords{planetary systems - planetary systems: formation}

\maketitle

\section{Introduction}
\label{sec:introduction}

Determining the planetary internal structure is the first important goal in characterizing planets and can provide important constraints
on their formation and habitability. 
In current planet
formation models, the composition of a planet strongly depends on its formation location, or more generally, on the location in the disk where
the planet has accreted the planetesimals constituting its core (see, e.g., Bond et al., 2010, Thiabaud et al. 2014, 2015a,b) and the gas constituting
its envelope (e.g., \"Oberg et al., 2011, Madhusudhan et al., 2014, Thiabaud et al., 2014, 2015a,b).  In addition,
the presence of volatiles in large quantities in a planet sets constraints on the thermal structure of the protoplanetary disk in which the
planet has formed (see, e.g., Marboeuf et al. 2014a,b). Indeed, volatile ices sublimate at low temperatures (below 170 K)
and are believed to be present in only small quantities in the innermost regions of a protoplanetary disk. Planets with
large amounts of volatiles at close distances from their star  therefore represent a strong argument in favor of planetary migration
(see, e.g., Baruteau et al, 2014) or transport of planetesimals in protoplanetary disks.
The amount of water also has important implications for the habitability of planets. It is known that water is necessary
for life as we know it to exist. However, too much water in a planet
is likely to prevent habitability by suppressing the C-cycle and its temperature stabilizing effect (Abott et al. 2012, Alibert, 2014, Kitzmann et al. 2015).

Deriving the internal composition of a planet from its mass and radius
is a highly degenerated problem, however (Seager et al., 2007, Sotin et al. 2007, Valencia et al. 2010, Rogers and Seager 2010). For example, a planet of a given density can be explained by an Earth-like core surrounded by a gas envelope,
but it can also be explained by a smaller core surrounded by a layer of ice and a thinner layer of gas. A numerical example of these planetary structures
is given in the following sections.

The goal of this paper is to compute the radius and the radius evolution of planets of different composition, in particular of different volatile and gas content.  {We emphasize that different
from previous studies, we explicitly exclude H and He from the class of 'volatiles' and consider them as 'gas'.} 
As we show below, changing the amount of volatiles in a planet not only modifies its radius, it also changes the mean heat capacity
of the planet and its gravitational energy by modifying the  mass repartition in the planetary interior. As a consequence, the amount 
of ices has an influence on the evolution of the planetary radius as a function of time. By measuring the radius distribution of  an ensemble
of planets of the same mass but at different ages, we show that it is possible to derive constraints on the cooling rate of the planets, and therefore on
their internal composition. This method can only be used for core-dominated planets because for gas giants the energy of the core is negligible
compared to the total energy of the planet. Finally, we would like to point out that our results strongly depend on the thermodynamics of
ices, silicates, and iron at very high pressure and temperature. As a consequence, it is more than likely that the evolution sequence of the different
types of planets we present is  modified before observational data are available to apply this model. Our calculations are therefore to be taken  as an illustration of the possible determination of ice fraction and not as clear results.
The paper is organized as follows: we present our internal structure model in Sect. 2.  {We then consider planets without any gas envelope,
for example, as a result of strong evaporation, and show in Sect. \ref{case_nogas} that their volatile content can be derived if their radius, mass, and the
composition of their central star are known}. We then concentrate in Sect. \ref{evolution} on planets with a gas envelope and describe our method of
deriving the evolution of a planet. In Sect. \ref{results}
we compute the distribution of radii of planets with and without
gas envelopes
at different epochs and compare them.  {We provide in Sect. \ref{examples}  some practical
example on how to use transit observations in the framework of PLATO to constrain the volatile content of extrasolar planets,}
 and we finally describe the limitations of our model and draw
some conclusions
 in Sect. \ref{conclusion}.

\section{Interior structure model}
\label{model}

 For the models we present here we assumed a simplified planetary composition, 
 following the approach originally developed by Sotin et al. (2007) that was also used in Alibert (2014). We made the following assumptions regarding the planetary structure:
\begin{itemize}
\item The only volatile specie we consider here is water ice.
\item The refractory material is made of Mg, Si, and Fe (and O). The Mg/Si and Fe/Si ratio are assumed for all planets to be equal to 1.131 and 0.986, respectively, except
for 55 Cnc e (see Sect. \ref{case_nogas}). These values, which are close
to solar, allow reproducing the mass and radius of Earth (see Sotin et al. 2007 for a discussion on the elemental composition of planets).
\item The Mg number (fraction of Mg in silicates, defined as Mg/(Mg+Fe) in the silicates) is equal to 0.9, again close to the Earth value (see Sotin et al, 2007).
\end{itemize}

We computed the internal structure of planets  that are assumed to be fully differentiated and that consist in four layers:
\begin{itemize}
\item a core,
\item a silicate mantle,
\item an icy mantle, and
\item a gas envelope consisting of H and He.
\end{itemize}
We did not consider the distinction between inner and outer mantle or a possible layer of liquid water at the surface in these models.
An icy mantle translates into high pressures at the boundary
of the icy to the silicate mantle, therefore no low-pressure silicate phase (e.g., olivine) can
exist because the transition between low- and high-pressure silicates occurs at a pressure of about 20 GPa.  A layer of liquid water
is possible if the gas envelope is very thin. However, as shown in Alibert (2014), the thickness and mass of a liquid water layer is negligible compared to the 
total radius and mass of planets. In addition, the planets we consider here have a gas envelope of at least a few percent in mass. The pressure at the interface of the gas to the solid planet\footnote{We refer to the three innermost layers (iron core, mantle, and volatile layer) as the 'solid planet'.}
 is therefore higher than the pressure at which ice VII appears (of about 2 GPa). 

\subsection{Structure of the solid planet}
\label{solidplanet}

To compute the temperature, density, and pressure in the solid planet, we solve the standard internal structure equations 
\begin{equation}
{d r \over d m } = { 1 \over 4 \pi r^2 \rho }
,\end{equation}
\begin{equation}
{d P \over d m } = { g \over 4 \pi r^2 }
,\end{equation}
and
\begin{equation}
{d T \over d m } = { g \nabla_{\rm ad} \over 4 \pi r^2}
,\end{equation}
where $P$ is the pressure, $r$ the radius, $m$ the mass interior to radius $r$, $g$ the gravity, $\rho$ and $\nabla_{\rm ad}$ the density
and adiabatic gradient given by the equation of state (see below), and $T$ the temperature. 
The equations were solved using the mass as an independent variable for each layer separately. The results provide the  structure of the solid planet, including its radius.  

The temperature profile in the different layers of the solid planet follows an adiabat. In contrast to other works (e.g., Sotin et al, 2007, Grasset et al. 2010, Alibert 2014), we did not
assume any temperature discontinuity at the transition between two layers. Given the uncertainties in the equations of state
(EOS) we used, we consider  this a reasonable assumption.
Adding temperature jumps would also not change our results qualitatively. As has been shown in different publications (e.g., Sotin et al. 2007,  Seager et al. 2007, 
Grasset et al. 2010, Valencia et al. 2010), the  radius of the solid planets at  a given mass hardly depends on the thermal profile.  

The boundary conditions for the solid planet interior model are as follows: we specified a surface temperature and pressure 
(in praxis these are given by the result of the planetary envelope model, see Sect. \ref{gasmodel}) and the planetary composition (in particular the volatile fraction). We recall that we consider that the only volatile species that can be present in a planet - more precisely, the only volatile species that can have a significant effect on the mass
and radius of the solid planet - is water ice. For that reason, we use 'ice' and 'volatile' without any distinction. We also recall that in contrast
to other studies, H and He are not counted as 'volatiles', but as 'gas'.  Finally, the thermal and gravitational energy
 of the solid planet were also computed as a result of the planetary interior structure, the specific heat being computed as a function of the local temperature,
 pressure, and composition - iron, silicate, or ice - as explained in  Sect. \ref{heatcapacity}.
  
\subsection{Equations of state}

Deriving the internal structure requires specifying the EOS, the adiabatic gradient and, to compute the thermal
energy, the heat capacity. The equations giving the pressure as a function of temperature and  density are similar to those used in Alibert (2014) and 
are reproduced here for the sake of completion. We refer to this paper and to Sotin et al. (2007) for more details and a justification of the use  of these EOS.  An in-depth introduction to these EOS is also presented in Poirier (2000).

We considered four different components
for the silicate layer, namely MgO , MgSiO$_3$, FeO, and FeSiO$_3$. The relative abundance of these different species in the
silicate mantle can be computed knowing the Mg/Si and Fe/Si ratios
and the Mg number. For the case considered here (values close to solar), the  fractions
of these species are 18.38 \%,  71.61 \%, 2.04 \%, and 7.95 \%,
respectively. For each of the components of the silicate mantle
and for the icy layer,
the EOS is given by the Mie-Gruneisen-Debye formulation (see Poirier 2000):
\begin{equation}
P = P(\rho,T_0) + \Delta P
,\end{equation}
\begin{equation}
\Delta P = \gamma \rho \left( E(T)-E(T_0) \right)
,\end{equation}

\begin{eqnarray}
\begin{array}{l}
 P(\rho,T_0) = {3 \over 2} K_0 \left[ \left( {\rho \over \rho_0} \right)^{7/3} -  \left( {\rho \over \rho_0} \right)^{5/3} \right] \\
\qquad  \qquad \qquad \qquad \times \left( 1 - {3 \over 4} \left( 4-{K_0}^\prime \right) \left[ \left( {\rho \over \rho_0 }\right)^{2/3} -1 \right] \right)
\end{array}
,\end{eqnarray}
and
\begin{equation}
E = { 9 n \over \mmol} P \left( { T \over \tetad  } \right)^3 \int_0^{\tetad / T} {x^3 e^x \over \left( e^x - 1 \right) } dx
,\end{equation}
where $n$ is the number of atoms in the considered compound.
The Debye temperature $\tetad$ is given by 
\begin{equation}
\tetad = \theta_{\rm D,0} \left( \rho \over \rho_0 \right)^\gamma
,\end{equation}
and $\gamma$ is given by $\gamma = \gamma_0 \left( \rho \over \rho_0 \right)^{-q}$. 

\begin{center}
\begin{table*}[ht]
\caption{Parameters of the Mie-Gruneisen-Debye  and Belonoshko EOS}
\begin{center}
\begin{tabular}{lccccccccc}
\hline\noalign{\smallskip}
 Specie & $\rho_0$ (g/cm$^3$) & $T_0$ (K) & $K_0$ (GPa) & $K_0^\prime$ (GPa/K) & $\theta_{D,0}$ (K) & $\gamma$ & $q$ & $\mmol$ (g) & $n$ \\
\noalign{\smallskip}
\hline\noalign{\smallskip}
MgO & 3.584 & 300 & 157 & 4.4 & 430 & 1.45 & 3 & 40.3 & 2 \\
MgSiO$_3$ & 4.108 & 300 & 263 & 3.9 & 1017 & 1.96 & 2.5 & 100.4 & 5\\
FeO & 5.864 & 300 & 157 & 4.4 & 430 & 1.45 & 3 & 80.1 & 2 \\
FeSiO$_3$ & 5.178 & 300 & 263 & 3.9 & 1017 & 1.96 & 2.5 & 131.9 & 5 \\
high-pressure ice & 1.46 & 300 & 23.9 & 4.2 & 1470 & 1.2  & 1.0 & 18 & 3 \\
Fe & 8.334 & 300 & 174 & 5.3 & - & 2.434 & 0.489 & 55.8 & - \\
\hline
\end{tabular}
\end{center}
\label{table_MGD}
\end{table*}
\end{center}

Finally, we used the EOS derived by Belonoshko (2010)
for pure Fe, which is similar to the Mie-Gruneisen-Debye EOS, but with a different thermal pressure term:
\begin{eqnarray}
\begin{array}{l}
P = {3 \over 2} K_{T,0}^0 \left[ \left( {\rho \over \rho_{0} } \right)^{7/3} -  \left( {\rho \over \rho_{0}} \right)^{5/3} \right] \\
\qquad  \qquad \qquad \qquad  \times \left(1 - {3 \over 4} \left( 4-{K_0}^\prime \right) \left[ \left( {\rho \over \rho_{0} }  \right)^{2/3} -1 \right] \right) \\
\qquad \qquad \qquad \qquad + 3 R \gamma (T-T_0) \times M/ \rho
\end{array}
,\end{eqnarray}
where the parameters are given in Table \ref{table_MGD}, and $\gamma$ has the same definition as for the Mie-Gruneisen-Debye EOS. 

The adiabatic gradient is computed as 
\begin{equation}
{d T \over d P} ={  \gamma T \over K_S } 
,\end{equation}
where $K_S$ is the adiabatic bulk modulus, which is related to the isothermal bulk modulus $K_T$ through
$K_S = \left( 1 + \gamma \alpha T \right) K_T$, $\alpha$ being the thermal expansion coefficient.
For silicate, $\alpha$ is given by
\begin{equation}
\alpha = \left( \alpha_0 + (T-T_0) a_1\right) \left( {\rho_0 \over \rho } \right)^{\delta_T} 
,\end{equation}
where $a_1$ is the derivative of the  expansion coefficient with respect to the temperature, and $\delta_T$ provides the dependance of $\alpha$ with respect to the density. 
$\tetad$ is the Debye temperature, which also depends on the density (see above). The numerical values of the different parameters are given in Table \ref{table_adiabat}
and are taken from Katsura et al. (2010) for the silicate mantle.

For the icy layer, we followed Valencia et al. (2006), the thermal coefficient being given by
\begin{equation}
\alpha = \left( \alpha_0 + a_1\right) \left( 1 + {K_0^\prime \over K_0} P  \right)^{- \delta_T} 
.\end{equation}
The values of the different constants are given in Table \ref{table_adiabat}.
For the iron core, the value of the thermal expansion coefficient was directly computed from the Belonoshko EOS (Belonoshko 2010).  

\begin{center}
\begin{table*}[ht]
\caption{Parameters used to compute the adiabatic gradient}
\begin{center}
\begin{tabular}{lccccccccc}
\hline\noalign{\smallskip}
 layer & $\delta_T$ & $a_1$ & $\alpha_0$ & $\gamma_0$ & $q$ & $\theta_0$ (K) & $\mmol$ (g) & $T_0$ (K) & $\rho_0$ (g/cm$^3$) \\
 \hline\noalign{\smallskip}
silicate mantle & 7.2 & $1.6 \times 10^{-8}$ & $2.56 \times 10^{-5}$  & 1.26 & 2.9 & 760 & 140.7 & 300 & 3.222 \\
icy layer & 1.1 & $1.56 \times 10^{-6}$ & $-4.2 \times 10^{-4}$ & 1.2 & 1 & 1470 & 18 & 300 & 1460 \\
\noalign{\smallskip}
\hline
\end{tabular}
\end{center}
\label{table_adiabat}
\end{table*}
\end{center}

\subsection{Heat capacity}
\label{heatcapacity}

The heat capacity was computed as follows. 
For ice, we used $C_v = C_{\rm max} D({\tetad \over T})$ (Stewart and Ahrens 2005), with D  the Debye integral 
$D (x) = {3 \over x^3} \int_0^x {{u^4 e^u \over (e^u -1)^2}} du$ and $ C_{\rm max} =  4600 $J/kg/K. 
For the iron core, $C_v$ was assumed to be constant equal to 40 J/mol/K (Wang et al. 2002). 
The heat capacity of the mantle was assumed to be constant and equal to 1200J/K/kg (Tackley et al. 2013).

\subsection{Gas envelope model}
\label{gasmodel}

The structure of the gas envelope was computed by solving the standard planetary evolution equations using the opacity of Freedman (2008) 
for solar composition and the Saumon-Chabrier EOS (Saumon, Chabrier, Van Horn, 1995). The irradiation from the central star was taken into account  using the two-stream
formalism of Guillot et al. (2010), modified according to Jin et al. (2014) and using an irradiation temperature equal to 250K. 
 The luminosity of the planet was a free parameter and was constant in the whole gas envelope.
Finally, we used the structure of the gas envelope to derive the pressure and temperature at the boundary of the solid planet
to the gas (see  Sect. \ref{solidplanet}), and
the transit radius, which is equal to the radius where the chord optical depth is equal to one. 
When we assumed a planet without a gas envelope (Sect. \ref{case_nogas}), the  pressure at the surface of the solid planet was assumed to be low and the  temperature was
assumed to be equal to the equilibrium temperature.

\section{Volatile content of gas-poor planets}
\label{case_nogas}

 {We first explore  in this section the special case of planets that do not contain any sizable gas envelope. In this case, the fraction of volatiles can be determined provided certain hypotheses are made on the refractory composition of the planet. When the Fe/Si and Mg/Si ratio in the planet are the same as that  of the central star (as supported, e.g., by the results of Thiabaud et al. 2015a,b), for instance, it is possible to
compute the planetary radius as a function of the fraction of volatiles. }

 {To illustrate this, we considered the close-in planet \object{55 Cnc e}. We took the parameters of this planet (mass, period, and equilibrium
temperature) from Gillon et al. (2012) and used the abundances quoted in Bond et al. (2010) for the 55 Cnc system. These parameters are summarized in Table \ref{table_parameters}.
The planet is located close enough to its central star and has a mass that is low enough to assume that any H/He envelope has been lost
(see Jin et al. 2014). It should be kept in mind, however, that this remains an assumption and depends, in particular, on the early history of the parent star. This quantity is poorly known. }

\begin{figure}
  \center
  \includegraphics[width=0.35\textheight]{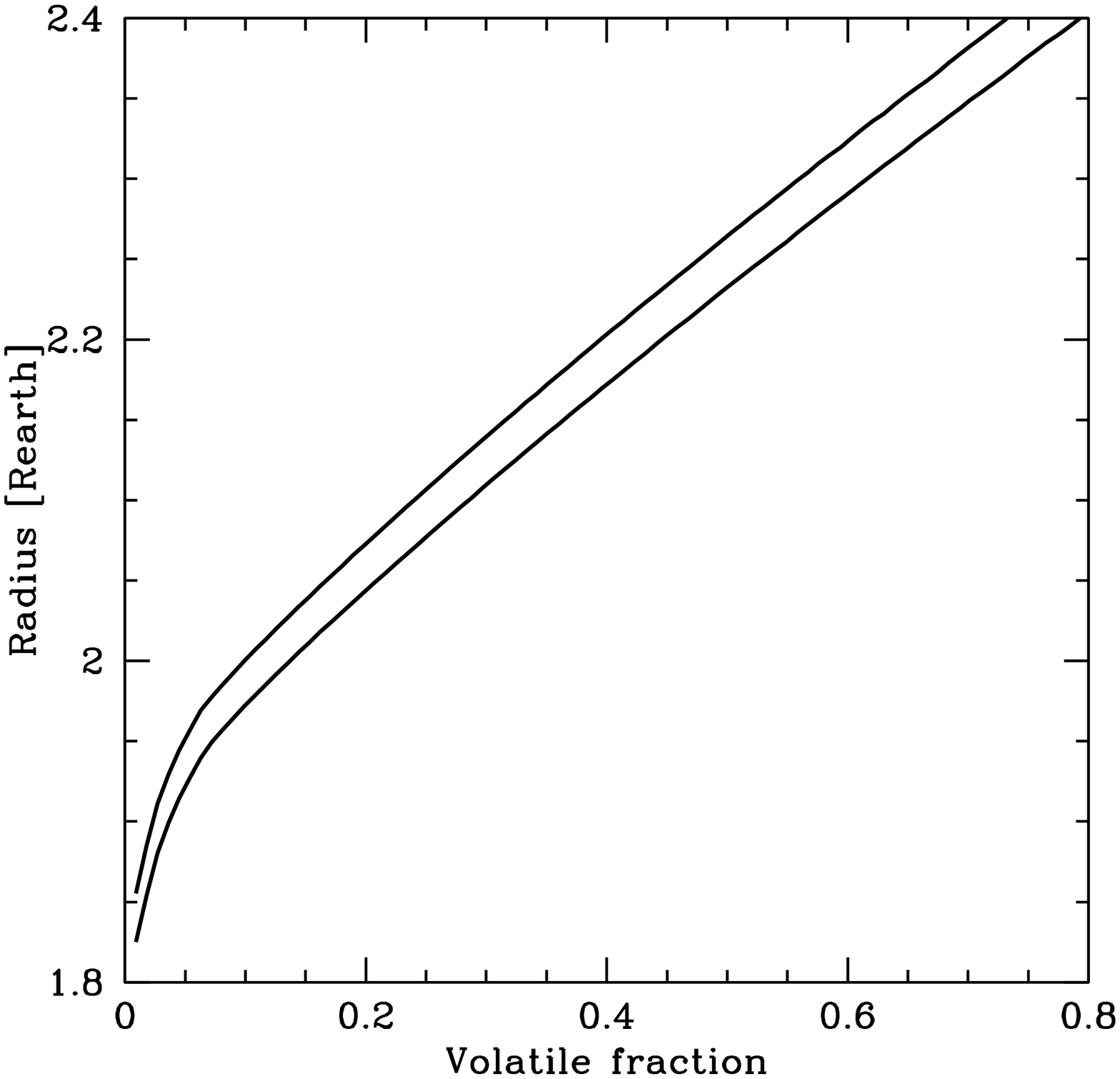}
  \caption{Planetary radius as a function of the fraction of volatiles for the minimum (dashed lines) and maximum (solid lines) masses of 55 Cnc e. The
  Fe/Si and Si/Mg ratio and the equilibrium temperature are taken from Table \ref{table_parameters}. }
  \label{nogas}
\end{figure}

 {Based on the observed composition of the stars, more precisely, their Fe/Si and Mg/Si ratios,
we computed the planetary radius as a function of the volatile fraction, assuming that $\fgas = 0$. By comparing the observed to the
computed radius, we can obtain a minimum and maximum value for $\fvol$ in the planet. The radius as a function of the volatile fraction is presented
in Fig. \ref{nogas}, the upper solid lines corresponding to the upper observed boundary of the planetary mass,
the lower dashed lines corresponding to the lower observed boundary of the planetary mass. 
From these calculations, we infer that the volatile fraction of 55 Cnc e lies between 19 \% and 56 \%, consistant with previous evaluations (see, e.g., Guillon et al. 2012).
Interestingly enough, the recent determination of 55 Cnc e radius in Demory et al. (2015) is much lower. In this case, a volatile fraction of 0 \% is compatible with the
mean density of the planet.}

\begin{center}
\begin{table*}[ht]
\caption{Assumed properties of 55 Cnc e}
\begin{center}
\begin{tabular}{lccccccc}
\hline\noalign{\smallskip}
 Planet & Minimum mass & Maximum mass & Minimum radius & Maximum radius & Fe/Si & Mg/Si & Equilibrium temperature \\
 \hline\noalign{\smallskip}
  55 Cnc e & 7.83 $\mearth$ & 8.35  $\mearth$ & 2.07  $\rearth$ & 2.27 $\rearth$ & 0.913 & 1.739 & 2400 K \\
\noalign{\smallskip}
\hline
\end{tabular}
\end{center}
\label{table_parameters}
\end{table*}
\end{center}

 {The key assumption in this calculation is that there is no sizable gas envelope, namely no gas layer that contributes significantly to the mass and radius
of the planet. Planets like this can be found close to the central star, where the evaporation process is assumed to be strong enough to remove any H/He that is acquired
during the formation (see, e.g., Jin et al. 2014, their Fig. 9).}

 {The observational determination of the volatile fraction in these planets is therefore possible, but it relies on a good understanding of evaporation models.
In particular, as shown in Guillon et al. (2012), a gas fraction of 0.1 \% would be enough to explain the observed radius of the planet without any volatiles.
The method outlined in this section moreover only applies to very close-in and  small planets.}

 { In the following, we concentrate on the other hand on planets located at a large distance from
the central star, where evaporation is assumed to be ineffective. In this case, the mass-radius relation is degenerate, and we need to rely on planetary evolution
calculations to set constraints on the fraction of volatiles.}

\section{Planetary evolution}
\label{evolution}
\subsection{Computing the time sequence}
\label{time_sequence}

The internal structure models presented above depend on the planetary luminosity $L$ as a free parameter.  We computed a set of 1000 planetary
structures for each planetary composition with luminosities from 100 $L_J$ to $10^{-5} L_J$, regularly spaced in log. The time evolution of a planet was then computed by assigning a time stamp to each structure
(labelled by its luminosity). To compute the time between two structures, we used the total energy conservation:
$$dt = {dE_{\rm tot} (t)  \over (L(t)-L_{\rm rad}(t)),}$$
where the luminosity coming from radioactive decay is proportional to the mass of silicates in the planet. The abundance of the different radioactive
elements ($^{40}$K, $^{238}$U, $^{232}$Th and $^{26}$Al) and their half-lifes were taken from Mordasini et al. (2012b).

The origin of time in the planetary evolution is determined as
the time when the radioactive decay rate is equal to 100 $\lj$;
this occurs very early in the planet evolution. As has been described in different studies (see Mordasini et al., 2012b and reference therein), the value of this starting time has no influence on the evolution of planets.

\subsection{Evolution tracks of planets with different water contents}

\subsubsection{Comparing planets with different volatile fractions and the same gas fractions}

\begin{figure}
  \center
  \includegraphics[width=0.35\textheight]{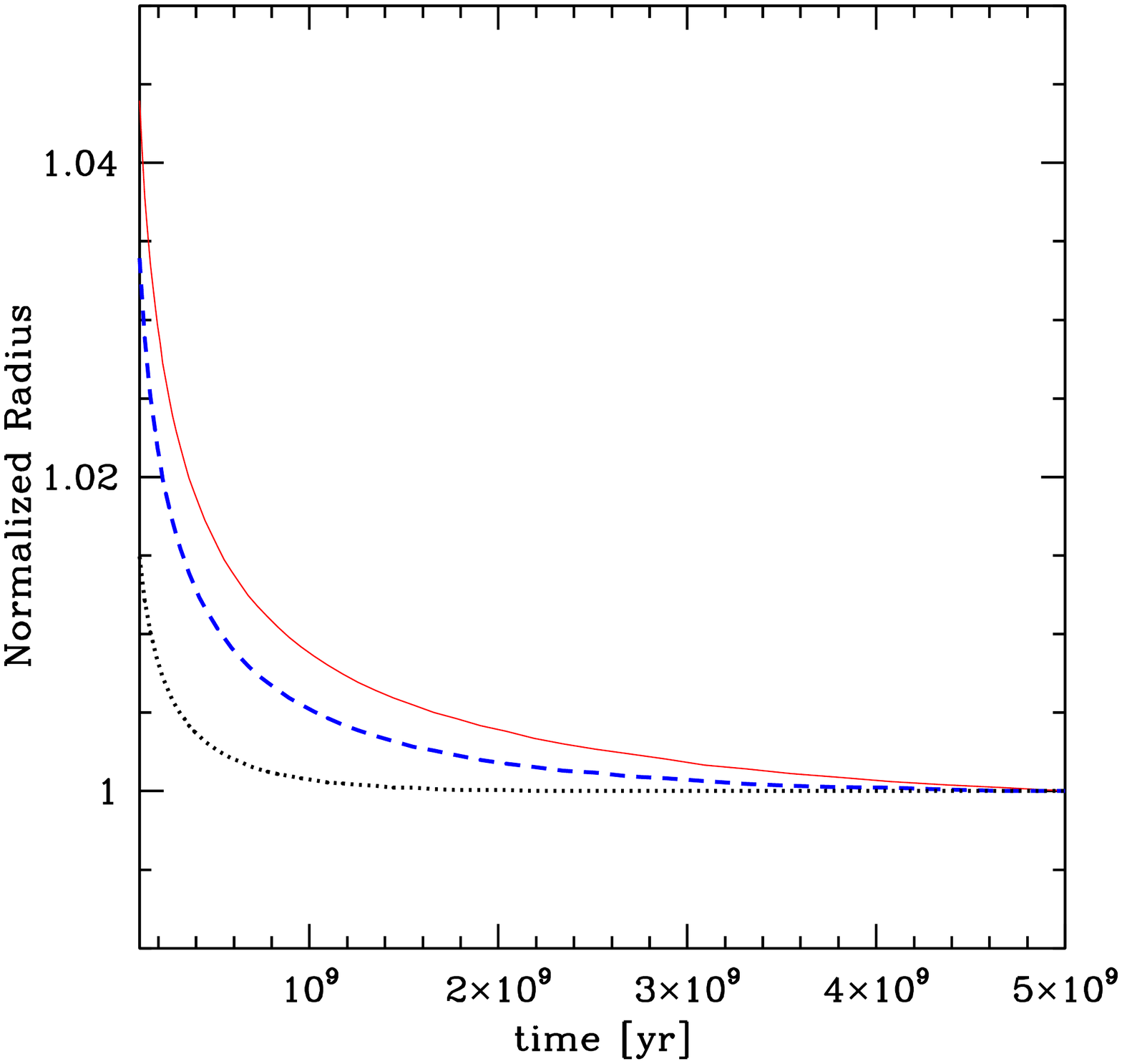}
  \caption{Normalized radius (defined as the ratio of the radius at a given time and the radius at 5 Gyr) as a function of time for three 12 $\mearth$ planet models. Solid line: volatile fraction $\fvol = 0$ and gas fraction $\fgas=0.1$. Dashed line:   volatile fraction $\fvol = 0.7$ and gas fraction $\fgas=0.1$. Dotted line: volatile fraction $\fvol = 0.7$ and gas fraction $\fgas = 0.04$. The first and third planets have by construction the same radius at 5 Gyr (3.23 $\rearth$), whereas the second one is larger (3.67 $\rearth$).}
  \label{evol1_new}
\end{figure}

 {As an example, we present in Fig. \ref{evol1_new} the evolution tracks of planets of 12 $\mearth$ with two different compositions. The first planet consists of 10 \% (all percentages
are given in mass) of gas,  its interior contains no volatiles, and its Fe/Mg/Si ratios are those quoted in Sect. \ref{model}. The second planet consists of 10\% of gas and its
interior contains 70 \% of ices. Considering such a high amount of volatiles is probably unrealistic, but it is important to note that depending on the formation scenario
of hot Neptunes, very different amounts of volatiles are expected. For a formation with migration (see, e.g., Alibert et al, 2013), we expect a very high percentage of volatiles),
whereas for in  \textit{\textup{situ}} formation (e.g., Chiang \& Laughlin 2013), hot Neptunes are predicted to be dry.}

The two considered planets do not have the same radius at any time, since the non-gaseous part of the planet has itself a different radius.
To compare the evolution of the two planets, we  plot the ratio of the radius at a given time to the radius at 5 Gyr. The two curves therefore join per definition
at 5 Gyr, where the actual planetary radii are 3.23 $\rearth$ and 3.67 $\rearth$ for the volatile-poor and volatile-rich planets,
respectively.
The plot shows that the two radii diverge at earlier epochs, the difference being about one percent at epochs earlier than 1 Gyr. 
A third planet is considered in the plot (dashed blue line) and is discussed in the next section.

 {In Fig. \ref{evol1_new} the two  planets have the same gas mass fraction, the difference in the evolution is  therefore only the result of
the increased fraction of volatiles (and consequently the decreased fraction of iron and silicates). As we show in Sect. \ref{phys_origin}, the more rapid evolution of the volatile-rich planet
is the result of differences in the energy of the two planets and  of their cooling rate. }

\subsubsection{Comparing planets with different volatile fractions and the same radius at 5 Gyr}

 {The two planets considered in Fig. \ref{evol1_new} do not have the same volatile fraction, but have the same gas fraction. As a consequence, these two planets
never have the same radius at any epoch. Observationally speaking, a planet of 12 $\mearth$  may be explained by one of the two models (the water-rich or the water-poor model),
but not by both models. In other words, the two models are very easy to distinguish based on the determination of the mass and radius of the planet. Comparing the evolution
of two planets for a given gas fraction is therefore interesting from the theoretical point of view (it allows demonstrating the effect of the volatile content), but is of less interest
for interpreting observations. The gas fraction of a planet is \textit{\textup{not}} an observable, but the planetary radius \textit{\textup{is}} an observable. It is therefore more logical,
from an observational point of view, to compare the evolution of two planets with the same mass and radius at a given epoch (5 Gyr in what follows). In this case, the two planets
(volatile-poor and volatile-rich planet) cannot have the same gas fraction, and the difference in the evolution will be the result of both the change in volatile fraction (as demonstrated
in the previous section) and of the change in gas fraction (as has been demonstrated in many papers, e.g., Nettelmann et al. 2010, Valencia et al. 2013, Lopez et al. 2013 for planets
in the mass range we consider here).} 

\begin{figure}
  \center
  \includegraphics[width=0.35\textheight]{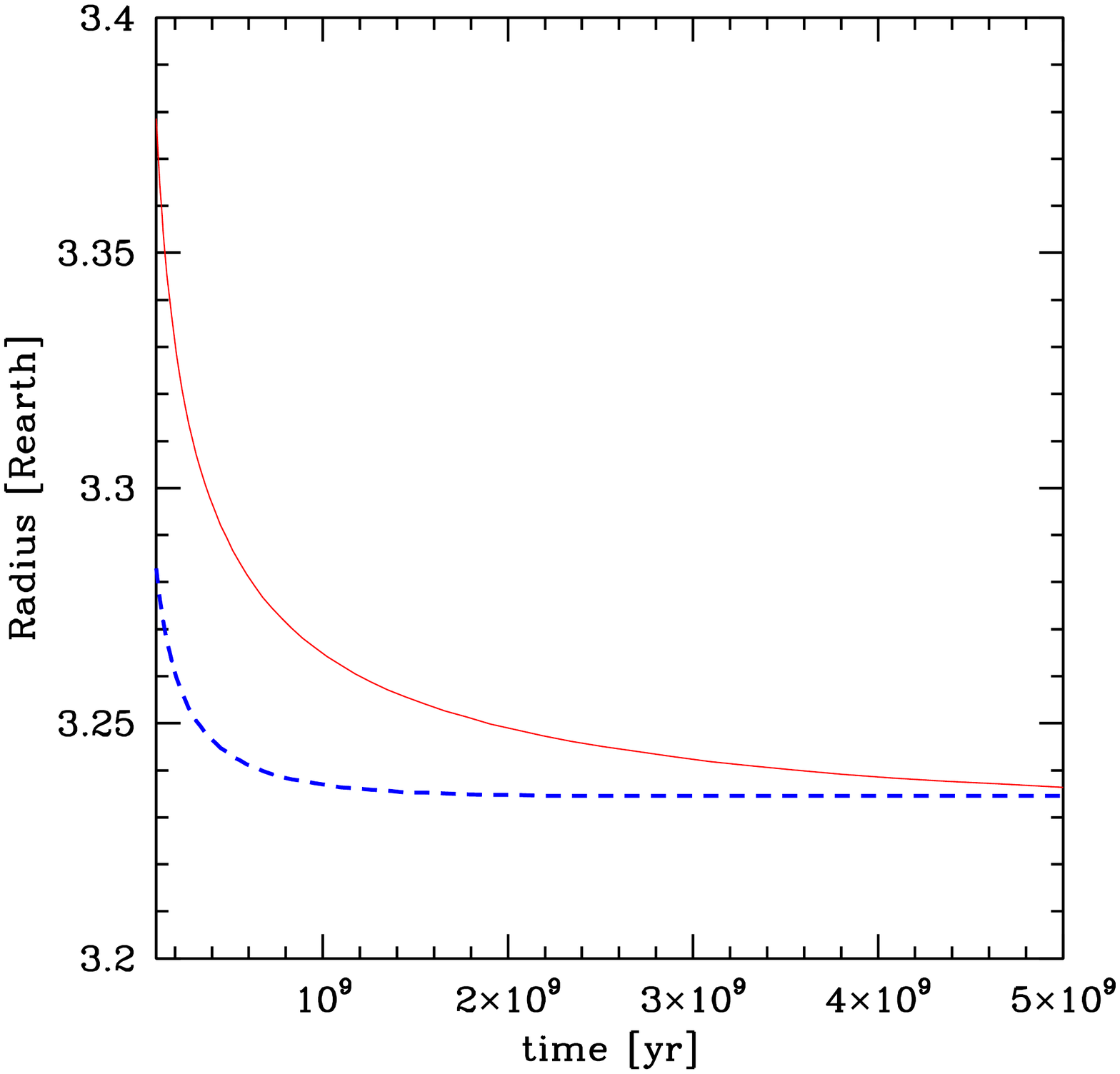}
  \caption{Radius as a function of time for two 12 $\mearth$ planet models. Solid line: volatile fraction $\fvol = 0$ and gas fraction $\fgas=0.1$. Dashed line:
  volatile fraction $\fvol = 0.7$ and gas fraction $\fgas=0.04$. }
  \label{evol1}
\end{figure}

 {To quantify the effect of changing the gas fraction on the
one hand  and the volatile fraction on the other, 
we considered a third  12 $\mearth$ planet with 70 \% of volatiles and a gas fraction of 4 \%. This gas fraction was adjusted to obtain a radius of 3.23 $\rearth$ at
5 Gyr, the same as that of the volatile poor planet. These two planets have, by construction, the same radius at 5 Gyr,  and the two internal structures therefore cannot be 
distinguished based on transit observations alone. Figure \ref{evol1} shows the radius evolution of the two planets as a function of time. The radii of the two 
planets diverge at earlier epochs, reaching $R = 3.37 \rearth$ for the volatile-poor planet and $3.28 \rearth$ for the volatile-rich planet at an age of 100 Myr. 
By comparing the three planets in Fig. \ref{evol1_new}, we see that the effect of the increased volatile fraction on the normalized radius at constant
gas fraction (red solid curve \textit{vs} blue dashed curve) is on the same order of magnitude as the effect of increased gas fraction at constant volatile fraction
(blue dashed curve \textit{vs} black dotted curve). We therefore conclude that both the gas and the volatile content of such planets are important for quantifying
their radius evolution.}

As a result of this difference in time evolution of the two planets with the same mass and same radius at 5 Gyr, it would be  possible to distinguish between the two internal structure by 
going back in time in the history of the planet. Since this is not possible, the alternative approach is to observe an ensemble of similar planets (with similar
total mass and distance to the central star, and orbiting similar stars) with different ages. By statistically comparing the  radius distributions at different ages, it is in principle possible
to statistically place constraints on the composition of planets. This approach relies on the assumption that the bulk composition of planets does not vary with time, which excludes
planets that are located too close to their star. In this case, evaporation modifies the gas content of the planet and complicates the problem. In the following, we therefore concentrate on
planets that are located far enough from the star to neglect evaporation. In addition, we neglect the possible accretion of matter by planets (e.g., in the
form of comets) on evolutionary timescales (after 100 Myr). This assumption is justified by two facts. First, in the case of Earth, the accretion of
comets was probably low (less than one percent of the Earth mass
because it would have to be smaller than the water inventory on Earth). Second, the planets
that we consider here contain high percentages of water, orders of magnitude larger than the estimated mass brought by comets on Earth.
 We note finally that this statistical approach also relies on the assumption that the planet formation process 
does not vary over billions of years, for example, as a result of changes in the composition of the interstellar medium on this timescale.

\subsubsection{Physical origin of the differential radius evolution}
\label{phys_origin}

 {The effect of the gas fraction on the radius evolution of low-mass planets has been discussed elsewhere (see, e.g., Nettelmann et al. 2010, Valencia et al. 2013, Lopez et al. 2013 among others).
We focus in this section on the differential evolution of the two planets with the same gas fraction, but different volatile fraction. 
The difference in radius evolution of  two planets with the same gas fraction but different volatile fraction is the result of two effects. }

\begin{figure}
  \center
  \includegraphics[width=0.35\textheight]{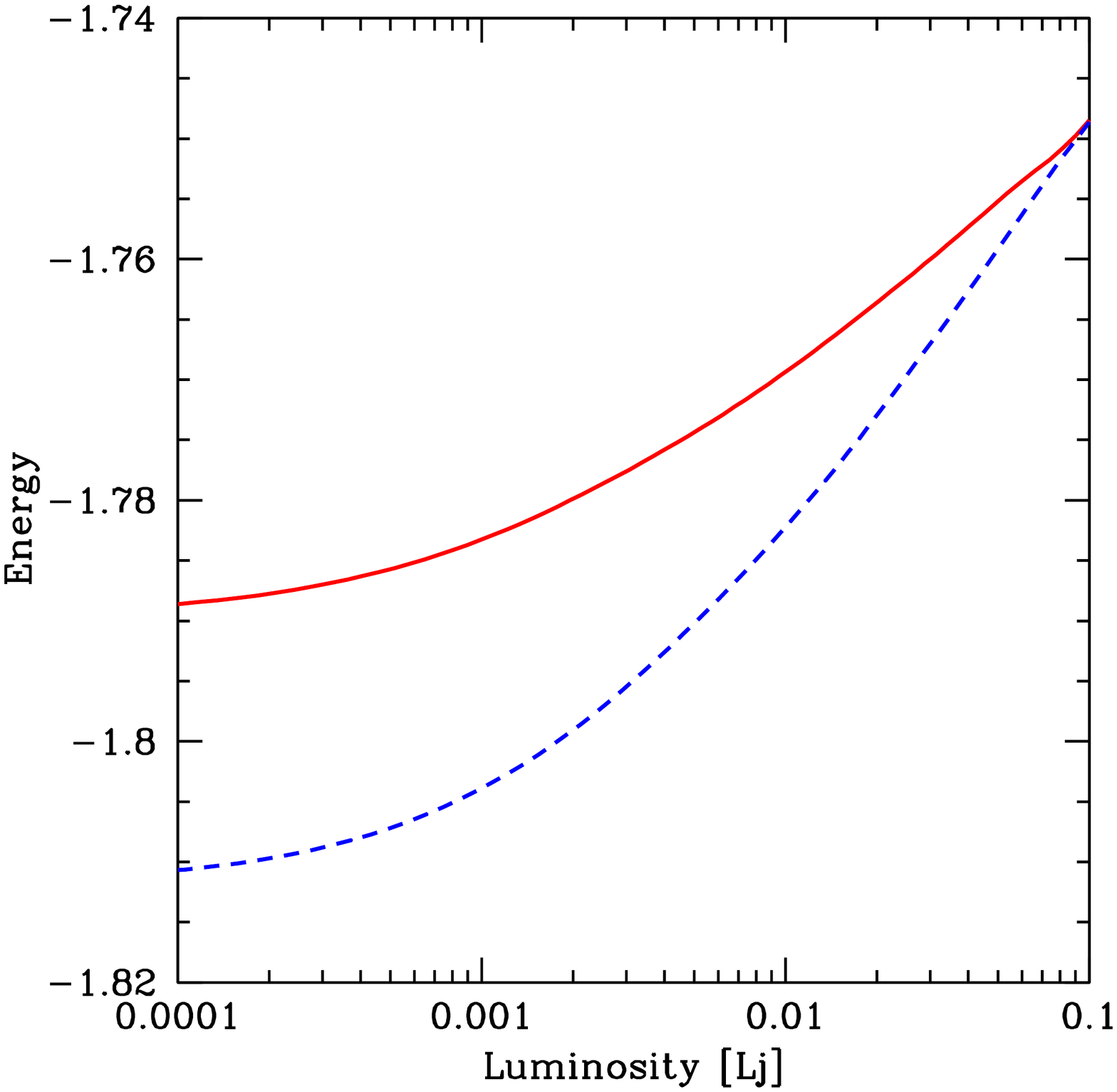}
  \caption{Total energy (in units of $10^{41}$ ergs) of a 12 $\mearth$ planet as a function of its luminosity. The solid red line represents a planet with $\fgas = 0.1$ and $\fvol = 0$, while the
  dashed blue line represents a planet with $\fgas = 0.1$ and $\fvol = 0.7$. To facilitate the comparison between the two curves, the energy of the volatile-rich
  planet has been multiplied by a constant factor equal to 0.39}
  \label{energy}
\end{figure}

The first effect is that the thermal and gravitational energy of planets depends on the volatile fraction: the thermal energy depends on the heat capacity, which is different for ices,
silicates, and iron, and the gravitational energy depends on the mass repartition in the planet. This is illustrated in Fig. \ref{energy}, which shows the total energy
of the two planets presented in Fig. \ref{evol1_new} (water-poor and water-rich planets, with the same amount of gas) as a function of their luminosity. 
As shown in Sect. \ref{time_sequence}, the time evolution of a planet depends on the relation
between the energy and the luminosity, which in turn depends on the internal composition of the planet.

A second effect is that the radioactive luminosity depends on the amount of silicates in the planet. Volatile-rich planets have fewer silicates and therefore less radioactive heating, which modifies their cooling.
We recall that the volatile fraction here is the fraction relative to the solid planet - a volatile-rich planet is therefore automatically a refractory-poor planet.

\section{Results}
\label{results}
\subsection{Radius distribution at different epochs}

\begin{figure}
  \center
  \includegraphics[width=0.35\textheight]{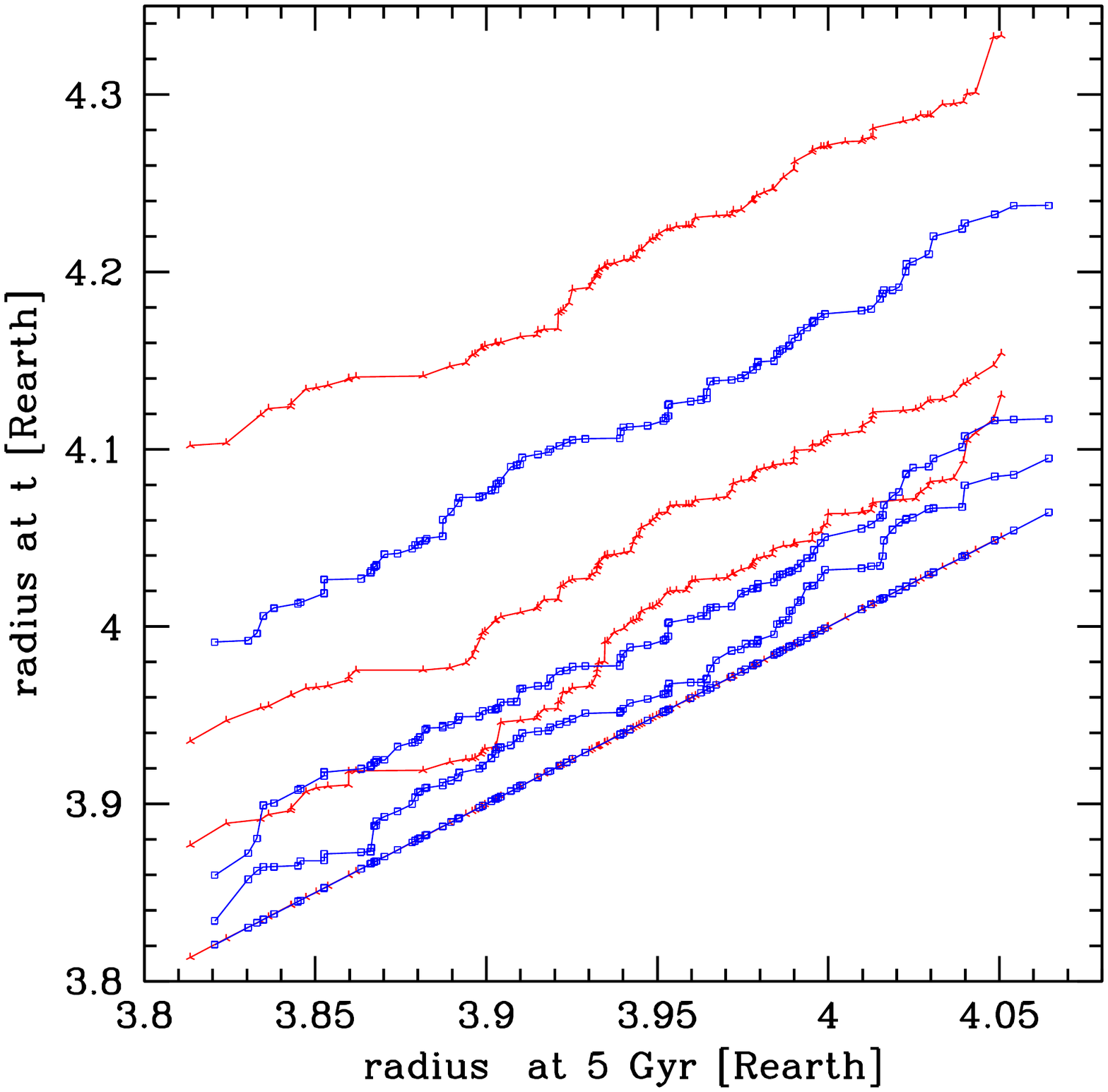}
  \caption{Quantile-quantile plot of the radius distribution at different ages. The reference age is taken to be 5 Gyr. The red three-branch stars represent the population of
  planets without volatiles and with a gas fraction of $20 \%$. The blue open squares show the population of planets with $70 \%$ volatiles in the interior and a mean
  value of $\fgas = 13.8 \%$ of gas. From top to bottom, the curves correspond to 100 Myr, 500 Myr, 1 Gyr, and 5 Gyr.}
  \label{qqplot}
\end{figure}

We now consider the evolution of the radius distribution at different epochs. To compute these distributions for different types of planets,
we computed a grid of $1.2 \times 10^6$ planetary structure for different gas fractions ($10^{-6}$, $10^{-5}$, $10^{-4}$, $10^{-3}$, $10^{-2}$, 0.03, 0.05, 0.06, 0.08,  0.1, 0.13, 0.17, 0.2, 0.3, 0.4, and 0.5),
different volatile fractions (0, 0.1, 0.3, 0.5, and 0.7) and different masses (1 to 15 $\mearth$ with 1 $\mearth$ step). For each of these cases, 1000 models were computed (each one corresponding to a different luminosity). When models for different parameters
were necessary, we interpolated in these tables to obtain the radius as a function of the luminosity.
To easily compare the distributions at different times, we used the radius distribution at 5 Gyr as a reference, and we plot the quantile of the radius distribution at a time $t$ as a function of the quantile
of the distribution at the reference time. These so-called quantile-quantile plots are interesting to visualize the evolution of the radius histogram as a function of time.
We present an example in Fig. \ref{qqplot} for two different cases:\begin{itemize}
\item case 1: the planetary mass ranges from 10 $\mearth$ to 15 $\mearth$ with a uniform distribution. The ice fraction is equal to 0 and the gas fraction to 20 \%.
\item case 2: same planetary masses as in case 1, the ice fraction is equal to 0.7, and the gas fraction is adjusted to have, on average, the same radius  as in case 1 ($\fgas = 13.8 \%$).
\end{itemize}
For each of these cases and for the different ages considered (100 Myr,  500 Myr, 1 Gyr, and 5 Gyr), the age was specified with an uncertainty of 10 \% 
and we added a random and uniformly distributed perturbation on the computed radius of 2 \%. These values were chosen to match the performance of PLATO 2.0
(see Rauer et al. 2014). We finally considered 100 planets for each case and each age.

In the two cases considered above, the mean gas fraction is different (20 \% in the volatile-poor case and 13.8 \% in the volatile-rich case). These two cases were
chosen to obtain two populations that \textit{\textup{at an age of 5 Gyr}} cannot be distinguished based on the measurement of the mass and radius. Considering
two populations with the same gas fraction would have resulted in two radius distributions that differed at 5 Gyr, therefore the two cases would have been easy to
distinguish based on the mass and radius measurement.

Figure \ref{qqplot} shows that the quantile-quantile plot moves toward the top of the figure when we consider past times. This is the result of the increase in planetary radius
when the luminosity increases. Moreover, comparing the red and blue symbols (the volatile-poor and volatile-rich planets), the quantile-quantile plot moves faster toward the top of
the diagram when the ice fraction is small. We recall that by construction, the radius distribution at 5 Gyr for the two populations is similar (the gas fraction was adjusted for that purpose).
The different behavior we observed in Fig. \ref{qqplot} is therefore the result of the difference in cooling and internal structure we explained in the previous section. 

\subsection{Comparing the distributions}

To facilitate comparing the distributions at different times and to assess the variability that is due to the uncertainty in age, mass, and observed radius,
we computed the vertical distance between the two radius cumulative distribution functions of the two populations (volatile-poor \textit{\textup{versus}} volatile-rich) at the same age\footnote{This distance is the  same as was used to compute Kolmogorov-Smirnov tests.}.
We then performed the same computation 1000 times and derived the distribution of these distances as a function of time. The
distribution of distances we find at 5 Gyr gives an order of magnitude of the intrinsic variability because in addition
to the uncertainties in radius and age, we observe planets with a range of mass.

\begin{figure}
 \center
\includegraphics[width=0.35\textheight]{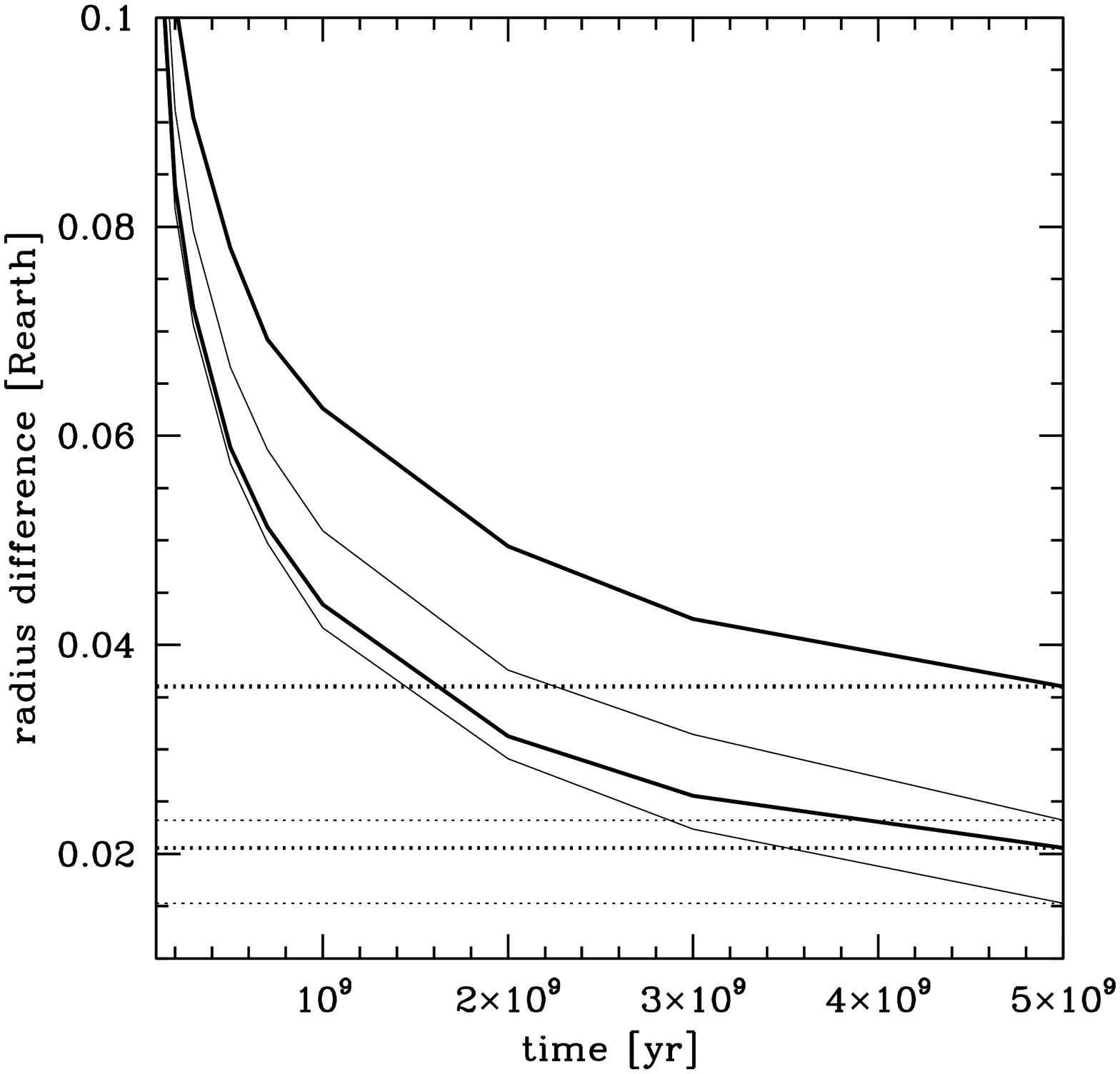}
\caption{Radius difference between the radius distribution of volatile-poor ($\fvol = 0$) and volatile-rich ($\fvol = 0.7$) planets at different
ages. The derivation of the radius difference is explained in the text, and the gas fraction of the two populations is adjusted so that the mean radius
of the two populations is the same at 5 Gyr, the gas fraction of the gas-poor population being 0.2. The heavy lines were computed assuming 100 planets are observed,
while the thin lines were computed assuming 500 planets are observed. The two solid lines give the 1 $\sigma$ range of the distance, while the
horizontal dotted lines are the 1 $\sigma$ variation of the mean distance at 5 Gyr. These latter values give the variability in distance that is caused by the uncertainties in the measurements and the variability of the planetary parameters. }
  \label{distance}
\end{figure}

As Fig. \ref{distance} shows, the two models (volatile-poor \textit{\textup{versus}} volatile-rich) can be distinguished for ages below $\sim$1.5 Gyr when 100 planets are observed at each age.
When the number of planets is increased to 500, the two models can be distinguished at later ages ($\sim 3$ Gyr), or stronger constraints can be set on the
fraction of volatiles in the interior of planets.

\subsection{Difference in radius distribution for different planetary types}

\begin{figure}
 \center
  \includegraphics[width=0.35\textheight]{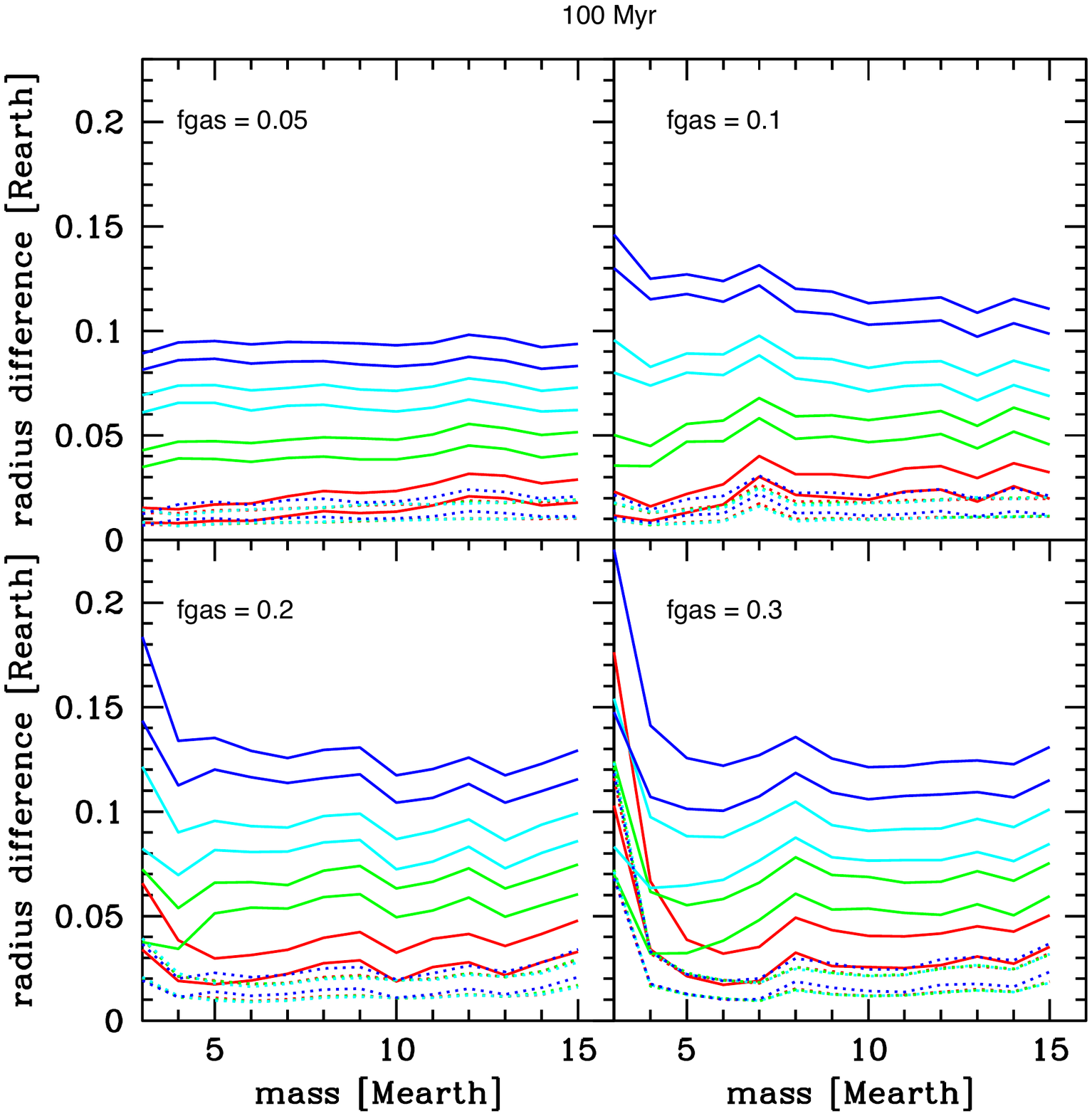}
  \caption{Distance between the radius distribution of volatile-poor  and volatile-rich planets as a function of their mass (horizontal axis) and volatile fraction (color: blue shows a
  70 \% volatile fraction, cyan 50 \%, green 30 \%, and red 10 \%). The two solid lines give the 1 $\sigma$ distribution of the distance obtained after running the same calculation 
  500 times. The dotted line shows the distance between the radius distribution at 5 Gyr (a measure of the intrinsic variability of the radius distribution -
also responsible for the noise evident in all curves), while the solid lines show
  the distance at 100 Myr. For this figure, we have considered 500 planets in each case.}
  \label{optimisation100Myr}
\end{figure}

\begin{figure}
  \center
   \includegraphics[width=0.35\textheight]{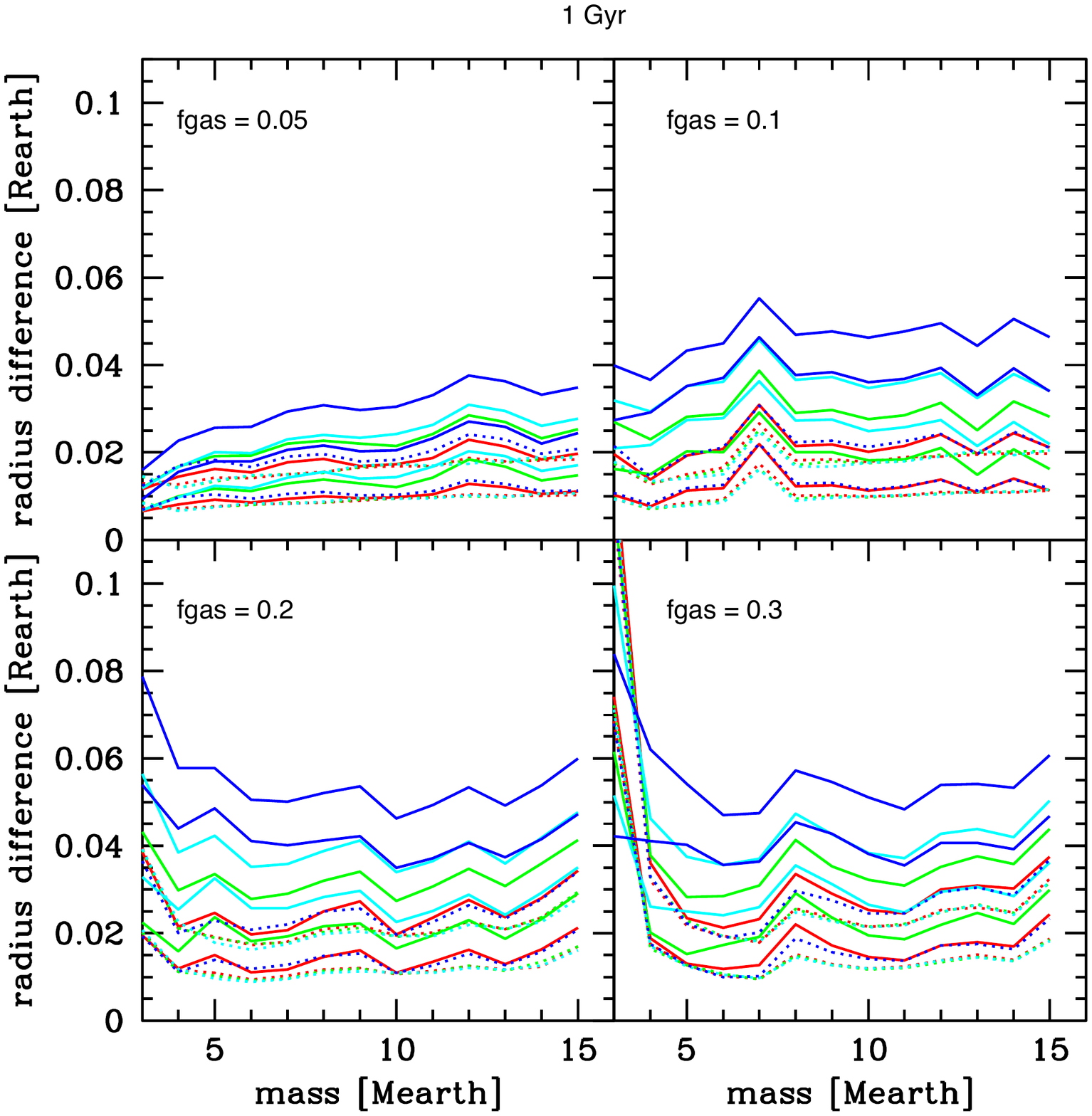}
  \caption{Same as Fig. \ref{optimisation100Myr}, but at 1 Gyr.}
  \label{optimisation1Gyr}
\end{figure}

To estimate which types of planets are better suited to estimate the volatile fraction, we computed the distance between two populations of planets
at different ages. The first population is a volatile-poor population, with different amounts of gas  ($\fgas =$ 0.05, 0.1, 0.2, and 0.3) and different masses (from 1 to 15 $\mearth$). 
The second population is a volatile-rich population, with an amount of gas that was adjusted so that the mean radius of the two populations at 5 Gyr was the same.
The gas fraction in the different planets is therefore equal to the value quoted in  Fig. \ref{optimisation100Myr} and Fig. \ref{optimisation1Gyr} \textit{\textup{only}} for the volatile-
poor planet, but  \textit{\textup{not}} for the other volatile fraction. We finally considered four volatile fractions for the second population, namely 10\%, 30\%, 50 \%, and 70 \%. We note that some of the planets considered in these two figures may not exist in nature because no
formation path leads to such an interior structure (see, e.g., Fortier et al. 2013, Alibert et al. 2013).

In each case (gas and volatile fraction), we computed the distance between the radius distributions of the first and second population as a function of the age.
We present in Figs.  \ref{optimisation100Myr} and \ref{optimisation1Gyr} the results for an age of 100 Myr and 1 Gyr. We repeated the simulation a few thousand
times to estimate the variability in the distance, and we plot in each panel of the two figures the $\pm$1-$\sigma$ interval of this distance. In the same panels, the
dashed lines represent the variability of the distance at 5 Gyr, which is a measure of the intrinsic variability of the radius distribution. The planets for which the volatile content
can be constrained are therefore those for which the solid lines lie above the two dashed lines of the same color.

Different conclusions can be drawn from these figures. The first is that the distance between the volatile-poor and volatile-rich radius distribution increases when considering younger
planets. This is a consequence of the effect observed in Fig. \ref{qqplot} where the red and blue curves move apart when considering younger objects.
The second conclusion is that the effect of the change in volatile content does not strongly depend on the planetary mass, except for very low mass planets (below 5 $\mearth$). Neither does the 
effect  depend strongly on the amount of gas, at least for $\fgas$ larger than 10 \%.

The effect of the volatile content is quite strong in all the cases considered here for planets in the 5-15 $\mearth$ mass range. Since in praxis the gas fraction of
planets is not a directly measurable quantity, planets in this mass range appear as the best option for quantifying their volatile fraction. Comparing statistically the radius distribution 
of such planets between 1 and 5 Gyr, or even better at 100 Myr and 5 Gyr, therefore appears to be the most promising way to place constraints on the volatile fraction, at least in the 
framework of the structure and evolution model presented here.

We considered in Figs. \ref{optimisation100Myr} and \ref{optimisation1Gyr} that 500 planets are observed for each planetary mass. This is of course a very large number and
may pose serious observational challenges. If we now consider that only 50 planets per mass bin are observed, we observe qualitatively the same results, although the dispersion
in the distances at different planetary masses, gas fraction, volatile fraction, and age increases.

\begin{figure}
  \center
     \includegraphics[width=0.35\textheight]{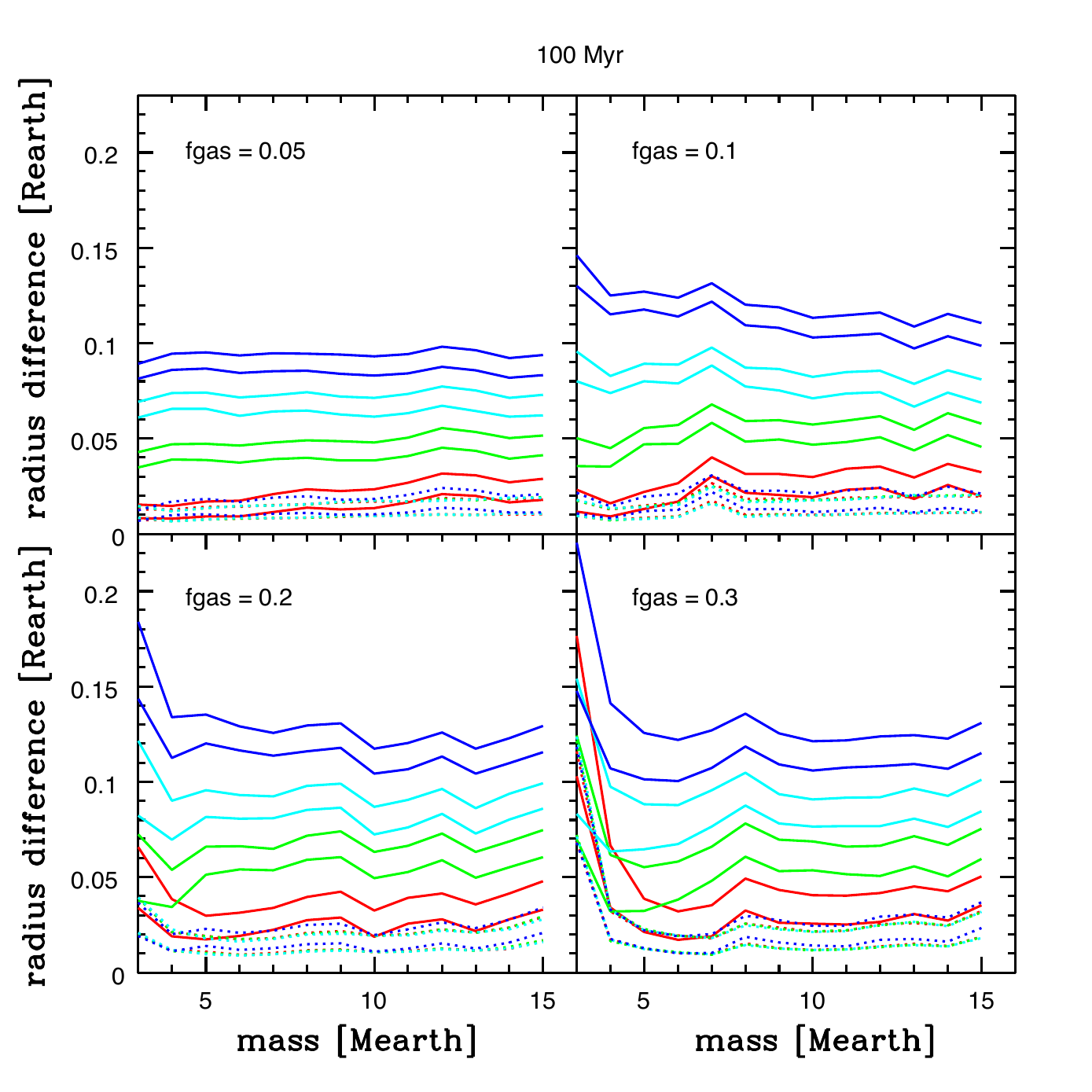}
  \caption{Same as Fig. \ref{optimisation100Myr}, but for 50 planets per mass bin.}
  \label{optimisation100Myr_50}
\end{figure}

\section{Practical examples}
\label{examples}

 {We provide two practical examples of how PLATO observations can be used to constrain the volatile fraction of an ensemble
of planets. We assume in the first example that PLATO has observed a set of 100 planets of 12 $\mearth$ (with 10 \% uncertainty) at 5 Gyr and another
set of similar planets (in terms of the distribution of volatiles and masses) at 100 Myr. These two ages are assumed to be accurate at 10 \%,
and all the radii are assumed to have an uncertainty of 2 \%. We moreover assume for this test that all planets have the same fraction
of volatiles, and we wish to demonstrate how we can constrain this volatile fraction.}

\begin{figure*}
  \center
  \includegraphics[width=0.6\textheight]{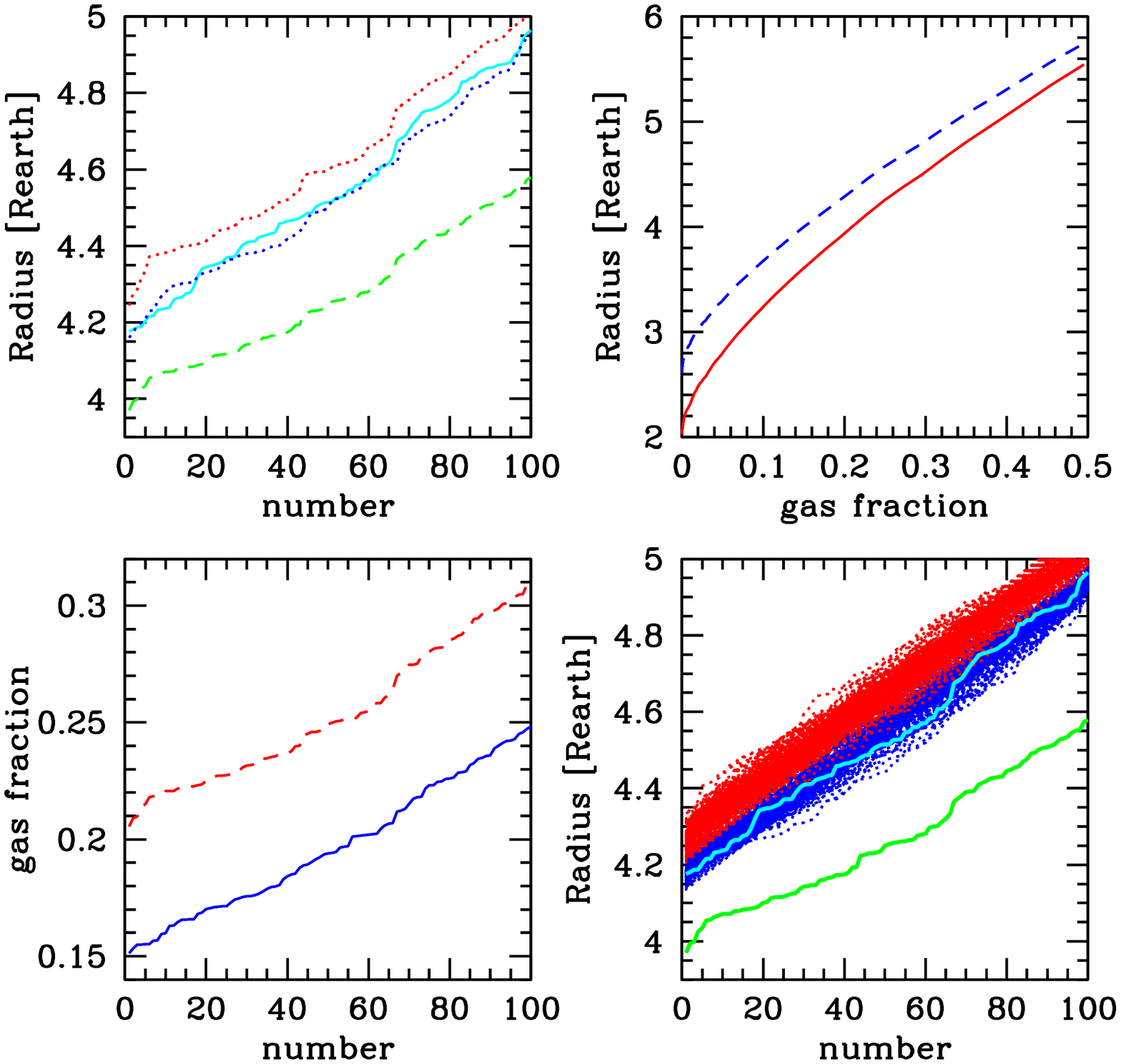}
  \caption{ {Top left: }Simulated cumulative distribution of two planetary samples at 5 Gyr (dashed line) and 100 Myr (solid line). The two
  dotted lines are the computed cumulative radius distribution at 100 Myr, assuming volatile-poor (upper curve) and volatile-rich (lower curve) planets.
  {Top right: }  Radius of a 12 $\mearth$ planet as a function of the gas fraction. The upper curve is computed assuming a volatile
  fraction $\fvol = 70 \%$, while the lower curve shows dry planets ($\fvol = 0$).
 {Bottom left:} Cumulative distribution of the gas fraction for volatile-rich planets (hypothesis 1) and volatile-poor planets (hypothesis 2).
 {Bottom right:}  Observed cumulative distributions of radius at 100 Myr (thick cyan line), at 5 Gyr (thick green line). The two sets of red and blue
lines are the computed cumulative radius distribution at 100 Myr  
  assuming volatile-rich planets (hypothesis 1, blue dotted curve) or volatile-poor planets (hypothesis 2, red dotted curve).  }
  \label{fig_all}
\end{figure*}

 {For this test, we computed  two sets of planetary structures, one at 5 Gyr and one at 100 Myr. The cumulative
radius distribution is presented in the top left panel of Fig. \ref{fig_all}. We then explore two hypothesis. The first is that the volatile
fraction of these planets is 0, the second that the volatile fraction of the planets is 70 \%.}

 {Under each of these hypothesis, we can compute the gas fraction of each of the observed planets of the sample observed at 5 Gyr.
For this, we used the curves plotted in  the top right panel
of Fig. \ref{fig_all}, which present the radius of a 12 $\mearth$ planet at 5 Gyr for different
values of the gas and volatile fraction. The two distributions of gas fraction are different for each hypothesis, the gas fraction of
volatile-rich planets (hypothesis 2) being smaller than the gas fraction of volatile-poor planets (hypothesis 1) because they have the same radius
at 5 Gyr. These two cumulative distributions are presented in  Fig. \ref{fig_all}, bottom left panel, and range from 15 \% to 25 \% for the volatile-rich planets, and
20 \% to 31 \% for the volatile-poor planets. }

 {With our evolution models we computed the cumulative distribution of radii at 100 Myr under the two hypotheses and compared
these two distributions to the observed one. The top left panel
of  Fig. \ref{fig_all} shows that the distribution of radii at 100 Myr under hypothesis 2 (planets
are volatile-rich) is a better fit than under hypothesis 1. To determine the effect of the different sources of variability in the radius distribution
(uncertainties on the mass, radius, and age, and intrinsic variation of the gas fraction of each planet), we repeated this calculation
100 times. In the bottom right panel of Fig. 10 we plot the cumulative distributions of radii at 100 Myr obtained using the same method. The figure clearly shows that the observed distribution at 100 Myr is statistically compatible with the distribution of volatile-rich planets (blue curves)
and not with the distribution of volatile-poor planets (red curves).}

\begin{figure*}
  \center
 \includegraphics[width=0.6\textheight]{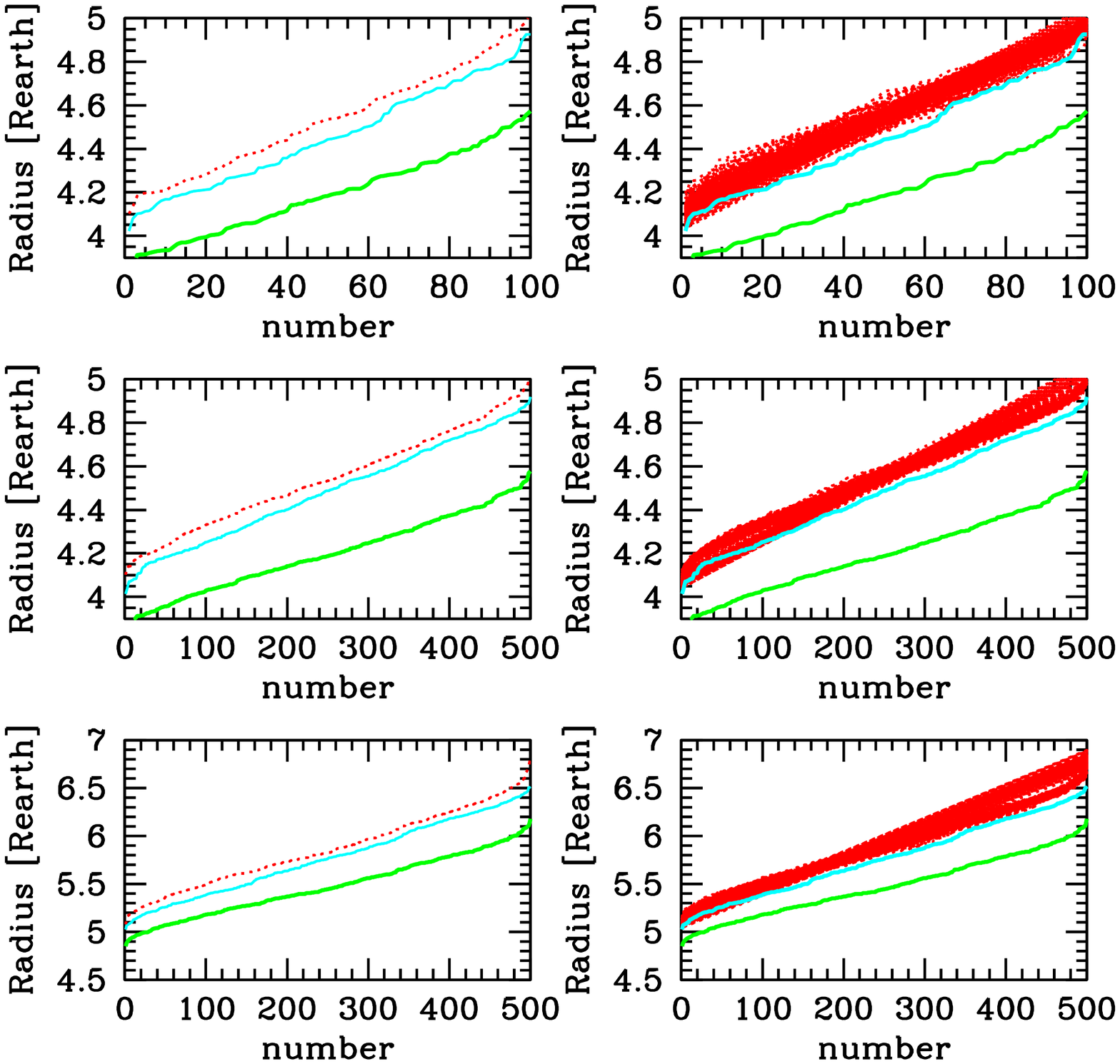}
  \caption{ {Upper left:} Simulated cumulative distribution of two planetary samples at 5 Gyr (lower solid green line) and 100 Myr (upper solid cyan line). The double 
  dotted line is the computed cumulative radius distribution at 100 Myr, assuming volatile-poor planets.
  {Top right: }   Observed cumulative distributions of radius at 100 Myr (upper solid cyan line) at 5 Gyr (lower solid green line). The set of red 
lines are the computed cumulative radius distribution at 100 Myr  assuming  volatile-poor planets.  
 {Middle line left and  right:} same as upper left and right panels, but assuming 500 planets are observed instead of 100 planets.
  {Lower  left and  right:} same as upper left and right panels, but assuming an equilibrium temperature of 1000 K.}
  \label{fig_test3}
\end{figure*}

 {This test is of course simplified because we assumed that all planets have the same volatile fraction. In reality,  this method still allows excluding certain hypotheses if there
is a diversity in the
volatile fraction of planets, provided enough planets are observed. To illustrate
this last point, we now consider a second test. We assume now that the considered planets (of measured mass equal to 12 $\mearth$ 
with 10 \% uncertainty) can have a range of volatile and gas fraction, taken to be 
$\fvol = 0.4$ to 0.7 and $\fgas$ = 0.15 to 0.25. We use the same method as in the first case and wish to determine whether it is possible to reject
the hypothesis that these planets are volatile-poor. Using the same method as above, we computed the cumulative distribution of planetary
radii at 100 Myr, assuming that they are volatile-poor. The result is presented in Fig. \ref{fig_test3}, assuming 100 planets or 500 planets are observed
in the upper and lower rows, respectively. The figure shows that the hypothesis that all planets are volatile-poor can be rejected from
observing 100 planets, which is even clearer from observing 500 planets.}

Finally, we also considered closer-in planets with an equilibrium temperature of 1000 K, still assuming that there is no change in the composition
of planets (e.g., because of evaporation) between 100 Myr and 5 Gyr. The results are presented in the last row of Fig. \ref{fig_test3}, which shows that the same conclusions
can also be drawn for high equilibrium temperatures.

To verify whether the effect presented above might be used to constrain the water fraction of planets, we ran a series of tests. The basis was a sample similar to the one presented in Marcy et al. (2014), assuming we now had 50 planets of similar mass whose radii are known with a precision of 5\%,
the mass with a precision of 50 \%. We also assumed that the age is known with 10 \% uncertainty, which is not the case in the Marcy et al. (2014) sample.  We then
ran five simulations, each time computing the radius distribution as we did in the first PLATO simulation presented above:
\begin{itemize}
\item considering this nominal sample,
\item decreasing the radius uncertainty to 2 \%, assuming this can be reached with the help of asteroseismology,
\item doubling the sample size,
\item decreasing the mass uncertainty to 20 \% for the
entire sample, and
\item taking all these improvements into account.
\end{itemize}
In all these cases except for the last, dry and wet cases are the same. This demonstrates that a precise determination of the mass together with a large enough sample are necessary to use the method we presented here.

\section{Discussion and conclusions}
\label{conclusion}

We have computed the evolution of planets in the mass range domain
of super-Earths to Neptunes for different compositions that were expressed in terms of gas fraction $\fgas$ and volatile
fraction $\fvol$. We recall here that we set the gas fraction equal to the fraction of H and He, excluding water or other volatiles.

{ We first considered planets without gas (located close enough to their star, see, e.g., Jin et al. 2014) and showed that the volatile fraction of a transiting
planet can be constrained by measuring the mass and radius. We illustrated this method on the example of 55 Cnc e, showing that
if gas
is absent from the planet, its possible volatile content ranges from 20 \% to 50 \%, using the radius determination of Gillon et al. (2012). 
This method is only applicable for very hot planets, however, and relies on the assumption
that the entire planetary gas envelope has been lost through evaporation.}

In a second part, we concentrated on planets located at a larger distance from their star, in a region where evaporation is ineffective.
We showed that although two planets could have the same radius and mass at a given time, in general they do not have the same radius at another
time. The difference is small, but large enough to be statistically measured by comparing two samples of planets of similar mass, one at 100 Myr or 1 Gyr, and one at 5 Gyr.
The possibility of measuring this difference depends of course critically on the precision of the measurements (we used the values expected for PLATO 2.0,
see Rauer et al. 2014) and on the number of planets and their mass.

From considering different planetary masses and compositions, we conclude that the planets that are best suited for placing constraints on the volatile
content are planets in the range of 5-15 $\mearth$ . In this case, observing about 100 planets in two age bins (below 1 Gyr and above 5 Gyr) would allow distinguishing
between volatile-poor planets and extremely volatile-rich planets (volatile fraction equal to 70 \%). Observing more planets, or at a higher precision, would translate into
a better precision of the volatile content.

The launch of PLATO is scheduled to occur in 2024. Before this date, CHEOPS (Broeg et al. 2013) and TESS (Ricker et al. 2010) will provide many
transit observations. These observations, however, are less usable in the framework of what is described in this paper, for two reasons. First, it is not clear if the age of
target stars will be known precisely enough. As was discussed above, knowing the age of planets (assumed to be the age of the star) is critical for using the method
we presented. Second, CHEOPS and TESS will mainly target planets located close to their star, for which it may be difficult to ignore the effect of evaporation
during the planet evolution.

The models we considered here are idealized, as are all models, in particular on at least two aspects. 
The first idealization is that we have implicitly assumed that planets  are homogeneous in their structure at least to some
extent. In particular, the tests illustrated
in Figs. \ref{optimisation100Myr} to \ref{optimisation100Myr_50} assume that the mass, gas fraction, and the volatile fraction of the planets is similar in the sample of planets (50 or 500). This is  only a small problem for the planetary mass because this is a measurable quantity, but it is more difficult for
the gas and volatile fraction. 

 {We have illustrated  the strategy to adopt in praxis to circumvent this problem.} Considering a population of planets of similar mass
with an observed radius distribution at 5 Gyr, different  models
can be built to explain the internal structure of the planets, for example, a first model that assumes no volatiles, and a second
that assumes 
a high percentage of volatiles. Based on these two assumptions, it is possible to determine the gas fraction distribution that allows reproducing the observed radius distribution at 5 Gyr.
For a population of planets in the same mass range but at much younger age, the radius distribution for similar planets can
be computed in the two models
considered above. By comparing these theoretical distributions with the observed distribution of young planets, some of the assumed models can be rejected and constraints 
on the volatile fraction are obtained.

The second idealization of the models is related to the internal structure  (including the gas envelope, the opacity, etc.). Our results depend,
as all internal structure models, on the assumed EOS of the different material, which enters the computation of the radius, but also on the gravitational and internal energy. 
The cooling of the planet and the contribution of the  envelope to the total radius also depends on the physics of the gas (opacity, mixing, and EOS).
{ It has been demonstrated that  the opacity effect on the radius is small (see Lopez and Forney 2014).}
Finally, we considered totally differentiated planets. In reality, it is likely that planets present a certain degree of mixing, which might particularly influence 
their gravitational energy. The energy of the planet is particularly important because it governs the time evolution of a planet and represents one of the key processes that are at the origin
of the different evolution of volatile-poor and volatile-rich planets. This aspect of the model will clearly evolve in the future when new theoretical and experimental determinations 
of the EOS of different material under planetary conditions will be available.  {It should be kept in mind, however, that since the method proposed here relies on the comparison
between different classes of planets (e.g., volatile-rich \textit{\textup{versus}} volatile-poor planets, see first example in Sect. \ref{examples}), some of these uncertainties may cancel out, at least in
part. This should be the case for the structure of the refractory and gaseous part, but obviously not for the volatile part of the planets when volatile-rich
and volatile-poor planets are compared.}

Despite the different assumptions and approximations in the model we used here, the general fact remains that
the energy of a planet depends on its internal structure (even for a given mass and radius). This dependence 
leads to different planetary evolutions for different compositions.
Future transit missions may be able to measure this effect. Transit observations such as will be performed by CHEOPS (Broeg et al. 2013), TESS (Ricker et al. 2010) or PLATO (Rauer et al. 2014), will be able to set statistical constraints on planetary composition
with this model, provided the stellar age is known with sufficient accuracy and enough planets can be observed with sufficient
mass and radius accuracy. 

\acknowledgements

This work was supported in part by the European Research Council under grant 239605. This work has been carried out within the frame of the National Centre for Competence in Research PlanetS supported by the Swiss National Science Foundation. The authors acknowledge the financial support of the SNSF. The author thanks the referee, Eric Lopez, for very constructive remarks that greatly improved the manuscript.

\end{document}